\documentclass[journal]{IEEEtran}

\usepackage{graphicx}   % Written by David Carlisle and Sebastian Rahtz

\usepackage{amsmath}   % From the American Mathematical Society
\usepackage{array}

\usepackage[latin1]{inputenc}
\begin{document}

\title{Online Pattern Recognition\\for the ALICE High Level Trigger}


%\author{V.~Lindenstruth$^1$, C.~Loizides$^{2,3}$, D.~Roehrich$^4$, B.~Skaali$^5$, T.~Steinbeck$^1$, R.~Stock$^2$,\\ H.~Tilsner$^1$, K.~Ullaland$^4$, A.~Vestb{\o}$^4$ and T.~Vik$^5$ for the ALICE Collaboration%
%\thanks{$^1$Kirchhoff Institut für Physik, Im Neuenheimer Feld 227, D-69120 Heidelberg, Germany}%
%\thanks{$^2$Institut für Kernphysik Frankfurt, August-Euler-Str. 6, D-60486 Frankfurt am Main, Germany}%
%\thanks{$^3$Corresponding author, email: loizides@ikf.uni-frankfurt.de}%
%\thanks{$^4$Department of Physics, University of Bergen, Allegaten 55, N-5007 Bergen, Norway}%
%\thanks{$^5$Department of Physics, University of Oslo, P.O.Box 1048 Blindern, N-0316 Oslo, Norway}%
%}%
\author{V.~Lindenstruth, C.~Loizides, D.~Röhrich, B.~Skaali, T.~Steinbeck, R.~Stock,\\ H.~Tilsner, K.~Ullaland, A.~Vestb{\o} and T.~Vik for the ALICE Collaboration%
\thanks{Manuscript received June, 15, 2003; revised September 30, 2003.}% 
\thanks{V.~Lindenstruth, T.~Steinbeck and H.~Tilsner are with Kirchhoff Institut für Physik, Im Neuenheimer Feld 227, D-69120 Heidelberg, Germany.}%
\thanks{C.~Loizides and R.~Stock are with Institut für Kernphysik Frankfurt, August-Euler-Str. 6, D-60486 Frankfurt am Main, Germany.}%
\thanks{D.~Röhrich, K.~Ullaland and A.~Vestb{\o} are with Department of Physics, University of Bergen, Allegaten 55, N-5007 Bergen, Norway.}%
\thanks{B.~Skaali and T.~Vik are with Department of Physics, University of Oslo, P.O.Box 1048 Blindern, N-0316 Oslo, Norway.}%
}%

\maketitle

% $Id: abstract.tex,v 1.6 2003/10/06 17:16:25 loizides Exp $

\begin{abstract}
%The ALICE detectors are expected to produce an overall data rate of up to 25 GByte/sec which will exceed the foreseen mass storage bandwidth of $\sim$1.25 GByte/sec by a factor of 20. 
The ALICE High Level Trigger has to process data online, in order to
select interesting (sub)events, or to compress data efficiently by
modeling techniques. %Data intensive repetitive local pattern
%recognition tasks will be done in custom hardware.
Focusing on the main data source, the Time Projection Chamber (TPC),
we present two pattern recognition methods under investigation: a
sequential approach (\emph{cluster finder} and \emph{track follower})
and an iterative approach (\emph{track candidate finder} and
\emph{cluster deconvoluter}). We show, that the former is suited for
pp and low multiplicity PbPb collisions, whereas the latter might be
applicable for high multiplicity PbPb collisions of dN/dy$\mathbf{>}$3000.
%if it turns out that more than charged particles would have 
%to be reconstructed inside the TPC. 
Based on the developed tracking schemes we show that using
modeling techniques a compression factor of around 10 might be
achievable. 
%Benchmarks of the former indicate, that we can reach the designed
%inspection rate for pp collisions of 1 kHz with more than 98\% and
%for low multiplicity PbPb collisions of 200 Hz with more than 90\%
%track detection efficiency. Though, the method will clearly fail for
%high multiplicity central PbPb collisions, if more than $\sim$8000
%charged particles are produced in the TPC. Here the latter, iterative
%approach will be applicable with an total efficiency of more than
%90\%.  

%In the following the different algorithms which has been implemented
%in the HLT framework together with the performance and status will be
%presented.  

%The system will consist of a farm of clustered SMP-nodes based on
%off-the-shelf PCs connected with a high bandwidth low latency
%network. Its nodes will be interfaced to the front-end electronics
%via optical fibers connecting to their internal PCI-bus, using a
%custom PCI Receiver Card (RORC). These boards provide a FPGA
%co-processor for data intensive repetitive task of the pattern
%recognition. 

%Most of the local pattern recognition will be done using the FPGA
%co-processor while the data is being transferred to the memory of the
%corresponding nodes. Algorithms for conventional cluster finding and
%local track finding based on a Circle Hough Transformation of the raw
%data are currently under development. Tests on prototypes are being
%done, using both the foreseen software for online data analysis and
%communication. Latest results will be shown. 
\end{abstract}

%\begin{keywords}
%\input{tex/keywords}
%\end{keywords}

% Will be ignored for other modes.
%\IEEEpeerreviewmaketitle

% $Id: introduction.tex,v 1.5 2003/10/06 17:16:25 loizides Exp $

\section{Introduction}
\label{introduction}

The ALICE experiment~\cite{alice} at the upcoming Large Hadron
Collider at CERN will investigate PbPb collisions at a center of mass
energy of about 5.5\,TeV per nucleon pair and pp collisions at
14\,TeV. Its main tracking detector, the Time Projection Chamber
(TPC), is read out by 557568 analog-to-digital channels (ADCs),
producing a data size of about 75\,MByte per event for central PbPb
collisions and around 0.5\,MByte for pp collisions at the highest
assumed multiplicities~\cite{alicetpc}.   

The event rate is limited by the bandwidth of the permanent storage
system. Without any further reduction or compression the ALICE TPC
detector can only take central PbPb events up to 20\,Hz and
min.\,bias\footnote{A minimum bias trigger selects events with as
  little as possible bias in respect to the nuclear cross section.} pp
events at a few 100\,Hz. Significantly higher rates are possible by
either selecting interesting (sub)events, or compressing data
efficiently by modeling techniques. Both requires pattern recognition
to be performed online. In order to process the detector information
at 10-25 GByte/sec, a massive parallel computing system is needed, the
High Level Trigger (HLT) system. 

\subsection{Functionality}
\label{functionality}

The HLT system is intended to reduce the data rate produced by the
detectors as far as possible to have reasonable taping costs. The key
component of the system is the ability to process the raw data
performing track pattern recognition in real time. Based on the
extracted information, clusters and tracks, data reduction can be done
in different ways: 
\begin{itemize}
\item {\bf Trigger}: Generation and application of a software trigger
  capable of selecting interesting events from the input data stream. 
\item {\bf Select}: Reduction in the size of the event data by
  selecting sub-events or region of interest. 
\item {\bf Compression}: Reduction in the size of the event data by
compression techniques.  
\end{itemize}

As such the HLT system will enable the ALICE TPC detector to run at a
rate up to 200\,Hz for heavy ion collisions, and up to 1\,kHz for pp
collisions. In order to increment the statistical significance of rare
processes, dedicated triggers can select candidate events or
sub-events. By analyzing tracking information from the different
detectors and (pre-)triggers  online, selective or partial readout of
the relevant detectors can be performed thus reducing the event rate.

The tasks of such a trigger are selections based upon the online
reconstructed track parameters of the particles, e.g. to select events
containing e$^+$e$^-$ candidates coming from quarkonium decay or to
select events containing high energy jets made out of collimated beams
of high $p_T$ particles~\cite{hltphysics}. In the case of low
multiplicity events such as for pp collisions, the online
reconstruction can be used to remove pile-up (superimposed) events
from the trigger event. 

\subsection{Architecture}
\label{architecture}

%The HLT system is located in the data flow after the front-end electronics and before the event building network. 
The HLT system receives data from the front-end electronics. A farm of
clustered SMP-nodes ($\sim$500 to 1000 nodes), based on off-the-shelf
PCs and connected with a high bandwidth, low latency network provide
the necessary computing power. The hierarchy of the farm has to be
adapted to both the parallelism in the data flow and to the complexity
of the pattern recognition. 

\begin{figure}[htb]
\begin{center}
\includegraphics[width=7cm]{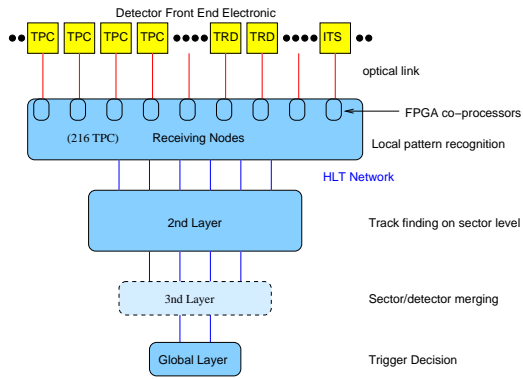}
\end{center}
\caption{Architecture of the HLT system.}
\label{hltarch}
\end{figure}

Fig.~\ref{hltarch} shows a sketch of the architecture of the
system. The TPC detector consists of 36 sectors, each sector being
divided into 6 sub-sectors. The data from each sub-sector are
transferred via an optical fiber from the detector front-end into 216
custom designed \emph{read-out receiver cards} (RORCs). Each receiver
node is interfaced to a RORC using its internal PCI bus. In addition
to the different communication interfaces, the RORCs provide a FPGA
co-processor for data intensive tasks of the pattern recognition and
enough external memory to store several dozen event fractions. A
hierarchical network interconnects all the receiver nodes.  

Each sector is processed in parallel, results are then merged in a
higher level. The first layer of nodes receives the data from the
detector and performs the pre-processing task, i.e. cluster and track
seeding on the sub-sector level. The next two levels of nodes exploit
the local neighborhood: track segment reconstruction on sector
level. Finally all local results are collected from the sectors or
from different detectors and combined on a global level: track segment
merging and final track fitting. 

The farm is designed to be completely fault tolerant avoiding all
single points of failure, except for the unique detector links. A
generic communication framework has been developed based on the
publisher-subscriber principle, which one allows to construct any
hierarchy of communication processing elements~\cite{pubsub}. 

% $Id: patternrecognition.tex,v 1.4 2003/10/06 17:16:25 loizides Exp $

%The expected data volume at the foreseen data rates of the different
%(sub-) detectors of ALICE exceeds the designed data acquisition (DAQ)
%bandwidth by a factor of 10 to 20. For a high event inspection rate
%the High Level Trigger (HLT) system has to process the data nearly in
%real-time doing online pattern recognition and simple online event
%reconstruction. The reconstructed information is used to either
%compress the data volume or to only store the ``interesting''
%fraction of the inspected events.  

%Around 500 to 1000 clustered SMP-nodes --possibly off-the-shelf
%personal computers-- will provide the necessary computing power for
%the whole HLT system. In addition, roughly 300 of these nodes are
%equipped with a Field Programmable Gate Array (FPGA) co-processor
%provided on a PCI bus card, which will be interfaced to the front-end
%electronics of the detector via optical fibers. Most of the local
%pattern recognition will be done using the FPGA co-processor while
%the data is being transferred to the memory of the corresponding
%nodes\cite{anders}.  

%Focusing on the Time Projection Chamber (TPC), we will present two
%methods for pattern recognition: the sequential and the iterative
%approach. Both methods are currently developed in software and partly
%also in hardware. A few details concerning the VHDL implementation
%will be given for the conventional cluster finder. 

\section{Online Pattern Recognition}
\label{onlinepatternrecognition}

%Concerning the TPC, which is the main data source of ALICE delivering
%up to 15 GB/sec, the HLT system has to perform online pattern
%recognition at an event rate of $\le$ 200 Hz for PbPb collisions and
%$\le$ 1 kHz for pp collisions. Up to $\sim$ 16000 charged particle
%tracks per PbPb event have to be reconstructed within a time budget
%of 5 ms and up to $\sim$ 1000 for pp in 1 ms.  
The main task of the HLT system is to reconstruct the complete event
information online. Concerning the TPC and the other tracking devices,
the particles should ideally follow helical trajectories due to the
solenoidal magnetic field of the L3 magnet, in which these detectors
are embedded. Thus, we mathematically describe a track by a helix with 
5(+1) parameters\footnote{To describe an arbitrary helix in 3 dimensions, 
one needs 7 continuous parameters and a handedness switch. For the 
special case of the ALICE geometry there are then 5 independent parameters 
plus the handedness switch.}.
%For track fitting only the 5 parameters are relevant as the handedness 
%will be deduced by the particle's path.
A TPC track is composed out of clusters. The pattern recognition task 
for the HLT system is to process the raw data in order to find clusters 
and to assign them to tracks thereby determining the helix track parameters 
using different fitting strategies.

For HLT tracking, we distinguish two different approaches: the
\emph{sequential feature extraction} and the \emph{iterative feature
extraction}.  

The sequential method, corresponding to the conventional way of event
reconstruction, first calculates the cluster centroids with a
\emph{Cluster Finder} and then uses a \emph{Track Follower} on these
space points to determine the track parameters. This approach is
applicable for lower occupancy like pp and low multiplicity PbPb
collisions. However, at larger multiplicities expected for PbPb at
LHC, clusters start to overlap and deconvolution becomes necessary in
order to achieve the desired tracking efficiencies.  

For that reason, the iterative method first searches for possible track
candidates using a suitable defined \emph{Track Candidate Finder} and
then assigns clusters to tracks using a \emph{Cluster Evaluator}
possibly deconvoluting overlapping clusters shared by different
tracks. 

For both methods, a helix fit on the assigned clusters finally
determines the track parameters.

%\subsection{Local pattern recognition}
%\label{localpattern}

%As emphasized in section \ref{architecture}, we try to do as much of
%the pattern recognition \emph{locally} defined by the readout
%granularity of the readout chambers to reduce data shipping and
%communication overhead in the HLT network system. %The TPC front-end
%electronic defines 216 data volumes, which are being read out over
%single fibers into the receiver nodes. 
In order to reduce data shipping and communicaton overhead within the
HLT, as much as possible of the \emph{local} pattern recognition will
be done on the RORC. We therefore intend to run the \emph{Cluster
Finder} or the \emph{Track Candidate Finder} directly on the FPGA
co-processor of the receiver nodes while reading out the data over the
fiber. In both cases the results, cluster centroids or track candidate
parameters, will be sent from the RORC to the memory of the host 
over the PCI bus. 

%Conventional cluster finding reconstructs positions of space points
%from raw data, which are interpreted as the crossing points between
%tracks and the center of padrows. The cluster centroids are
%calculated as the weighted charge mean in pad and time
%direction. Each padrow can processed independently, which is one of
%the main reasons to use a circuit for the parallel computation of the
%space points \cite{clusterfpga}.  

%To find track candidates, we use a variant of the \emph{Circle Hough
%Transformation}. The global detection of the track parameters is
%converted into a local peak detection in the parameter space. The
%transformation is directly applied on the raw data: Every charge
%value votes for all tracks it possibly belongs to. The intersection
%of these curves --seen as peaks in suitably defined histograms--
%define the track candidates. Again most of the computational effort
%can be highly parallelized in hardware \cite{dieter}.  

% $Id: clusterfinder.tex,v 1.8 2003/10/06 17:16:25 loizides Exp $

\subsection{Sequential Tracking Approach}
\label{seqtracker}

The classical approach of pattern recognition in the TPC is divided
into two sequential steps: Cluster finding and track finding. In the
first step the {\it Cluster Finder} reconstructs the cluster
centroids, which are interpreted as the three dimensional space points
produced by the traversing particles. The list of space points is then
passed to the {\it Track Follower}, which combines the clusters to
form track segments. A similar reconstruction chain has successfully
been used in the STAR L3 trigger~\cite{startrigger}, and thus has been
adapted  to the ALICE HLT framework.  

\subsubsection{The Cluster Finder}
\label{clusterfinder}

The input to the cluster finder is a list of above threshold timebin
sequences for each pad. The algorithm builds the clusters by matching
sequences on neighboring pads. In order to speed up the execution time
every calculation is performed {\it on-the-fly}; sequence centroid
calculation, sequence matching and deconvolution. Hence the loop over
sequences is done only once. Only two lists of sequences are stored at
every time: the current pad and the previous pad(s). For every new
sequence  the centroid position in the time direction is calculated by
the ADC weighted mean. The mean is then added to a current pad list,
and compared to the sequences in the previous. If a match is found,
the mean position in both pad and time is calculated and the cluster
list is updated. Every time a match is not found, the sequence is
regarded as a new cluster.  

In the case of overlapping clusters, a crude deconvolution scheme can
be performed\footnote{The deconvolution can be switched on/off by a
flag of the program}. In the time direction, overlapping sequences are
identified by local minima of the charge values within a sequence.
These sequences are separated by cutting at the position of the 
minimum in the time direction. The same approach is being used 
for the pad direction, where a cluster is cut if there is a local 
minimum of the pad charge values.

The algorithm is inherently local, as each padrow can processed
independently. This is one of the main reasons to use a circuit for
the parallel computation of the space points on the FPGA of the 
RORC~\cite{clusterfpga}.

\subsubsection{The Track Follower}
\label{trackfollower}

The tracking algorithm is based on {\it conformal mapping}. A space
point (x,y) is transformed in the following way: 
\begin{equation*}
x' = \frac{x-x_t}{r^{2}}
\end{equation*}
\begin{equation*}
y' = -\frac{y-y_t}{r^{2}}
\end{equation*}
\begin{equation}
r^2 = (x-x_t)^2 + (y-y_t)^2\,,
\end{equation}
where the reference point $(x_t,y_t)$ is a point on the trajectory of
the track. If the track is assumed to originate from the interaction
point, the reference point is replaced by the vertex coordinates. The
transformation has the property of transforming the circular
trajectories of the tracks into straight lines. Since then fitting
straight lines is easier and much faster than fitting circles (if we
neglect the changes in the weights of the points induced by conformal
mapping), the effect of the transformation is to speed up the track
fitting procedure.  

The track finding algorithm consists of a {\it follow-your-nose}
algorithm, where the tracks are built by including space points 
close to the fit~\cite{yepes96}. The tracks are initiated by building track
segments, and the search is starting at the outermost padrows. The
track segments are formed by linking space points, which are close in
space. When a certain number of space points have been linked together,
the points are fitted to straight lines in conformal space. These tracks
are then extended by searching for further clusters, which are close 
to the fit.  

\subsubsection{Track Merging}
\label{trackmerging}

Tracking can be done either locally on every sub-sector, on the sector
level or on the complete TPC. In the first two scenarios, the tracks
have to be merged across the detector boundaries. A simple and fast
track merging procedure has been implemented for the TPC. The
algorithm basically tries to match tracks which cross the detector
boundaries and whose difference in the helix parameters are below a
certain threshold. After the tracks have been merged, a final track
fit is performed in real space. 

%integral efficiency plots
\begin{figure}[thb]
\centering
\includegraphics[width=8cm]{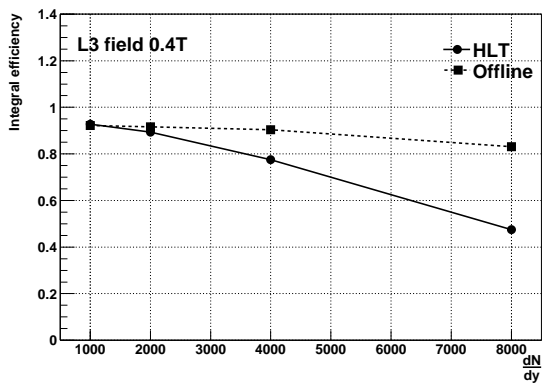}
\caption{Integral tracking efficiency for HLT online and ALIROOT offline
reconstruction as a function of different particle multiplicities 
for B=$0.4$T.}
\label{effint}
\end{figure}

\subsubsection{Tracking Performance}
\label{trackperformance}

The tracking performance has been studied and compared with the
offline TPC reconstruction chain. In the evaluation the following
quantities has been defined: 

\begin{itemize}
\item {\it Generated good track} -- A track which crosses at least
40\% of all padrows. In addition, it is required that half of the
innermost 10\% of the clusters are correctly assigned.
\item {\it Found good track} -- A track for which the number of assigned
clusters is at least 40\% of the total number of padrows. In addition,
the track should not have more than 10\% wrongly assigned clusters.
\item {\it Found fake track} -- A track which has sufficient amount
of clusters assigned, but more than 10\% wrongly assigned clusters.
\end{itemize}

The tracking efficiency is the ratio of the number of {\it found good
tracks}  to the number of {\it generated good tracks}. For comparison, 
the identical definitions have been used both for offline and HLT.  

Fig.\,\ref{effint} shows the comparison of the integral efficiency of
the HLT and offline reconstruction chains for different charged
particle multiplicities for a magnetic field of B=$0.4$T. We see that
up to dN/dy of 2000 the HLT efficiency is more than $90$\%, but for higher
multiplicities the HLT code becomes too inefficient to be used for
physics evaluation. In this regime other approaches have to be
applied. 

\subsubsection{Timing Performance}
The TPC analysis in HLT is divided into a hierarchy of processing
steps from cluster finding, track finding, track merging to track
fitting.  

\begin{figure}[hbt]
\centering
\includegraphics[height=5cm,width=7cm]{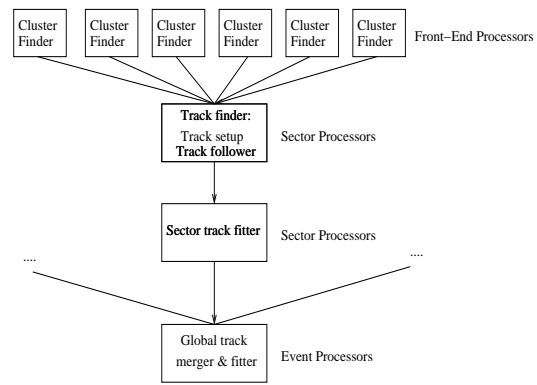}
\caption{HLT processing hierarchy for 1 TPC sector (= 6 sub-sectors)}
\label{hierarchy}
\end{figure}

Fig.\,\ref{hierarchy} shows the foreseen processing hierarchy for the
sequential approach. Cluster finding is done in parallel on each
Front-End Processor (FEP), whereas track finding and track fitting is
done sequentially on the sector level processors. The final TPC tracks
are obtained on the event processors, where the tracks are being
merged across the sector boundaries and a final track fit is performed
(compare to Fig.\,\ref{hltarch}). 

\begin{figure}[hbt]
\centering
\includegraphics[width=8cm]{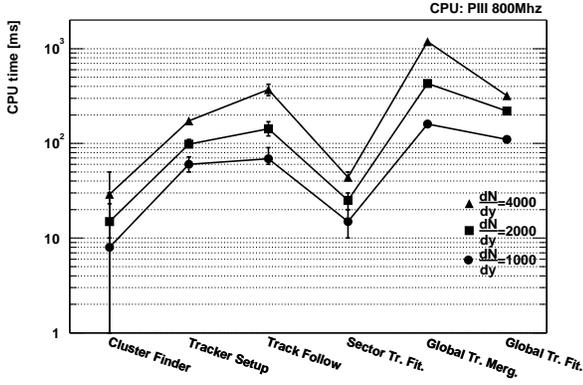}
\caption{Computing times measured on an P3 800\,MHz dual processor for 
different TPC occupancies and resolved with respect to the different 
processing steps.}
\label{timing}
\end{figure}

Fig.\,\ref{timing} shows the required computing time measured on a
standard reference PC\footnote{800\,MHz Twin Pentium III, ServerWorks
Chipset, 256\,kB L3 cache} corresponding to the different processing
steps for different particle multiplicities. The error bars denote the
standard deviation of processing time for the given event
ensemble. For particle multiplicity of dN/dy=4000, about 24 seconds
are required to process a complete event, or 4800 CPUs are required to
date for the TPC alone at an event rate of 200\,Hz\footnote{The estimate
ignores any communication and synchronization overhead in order to
operate the HLT system.}. 

\begin{table}
\begin{center}
\caption{Integral computing time comparison performance}
%\caption{Computing time comparison between HLT and Offline cluster finder and 
%track finder algorithms for reconstructing the whole TPC at dN/dy=4000.}
\label{tottime}
\begin{tabular}{|c|c|c|}
\hline 
dN/dy=4000 & \multicolumn{2}{|c|}{\bf CPU time (seconds)}\\
\cline{2-3} & HLT & Offline \\
\hline 
\hline
% {\bf Cluster finder} & 6 & 106\\
% {\bf Track finder}   & 18 & 58\\
{\bf Cluster finder} & 6 & 88\\
{\bf Track finder}   & 18 & 48\\
\hline
\end {tabular}
\end{center}
\end{table}

Table\,\ref{tottime} compares the CPU time needed to
reconstruct a TPC event of $dN/dy=4000$ for HLT and offline. 
In both cases, loading the data into memory was not included in the 
measurements\footnote{For offline, in addition 28 seconds are 
needed for data loading.}, in order to purely compare the two 
algorithms. For the overall performance of the HLT 
system, however, other factors as the transparent 
publisher-subscriber interface and network latencies 
become more important to allow an overall throughput 
with the expected rates. 
%For offline, loading the data into memory is also included in the
%measurement, while the HLT result only included the processing time as
%memory accesses are done completely transparent by the
%publisher-subscriber model\footnote{For offline 17 \% of the time is
%needed for data loading.}. 

% $Id: hough.tex,v 1.6 2003/10/06 17:16:25 loizides Exp $

\subsection{Iterative Tracking Approach}
\label{ittracking}

For large particle multiplicities clusters in the TPC start to
overlap, and deconvolution becomes necessary in order to achieve the
desired tracking efficiencies. The cluster shape is highly dependent
on the track parameters, and in particular on the track crossing
angles with the padrow and drift time. In order to properly
deconvolute the overlapping clusters, knowledge of the track
parameters that produced the clusters are necessary. For that purpose
the Hough transform is suited, as it can be applied directly on the
raw ADC data thus providing an estimate of the track parameters. Once
the track parameters are known, the clusters can be fit to the known
shape, and the cluster centroid can be correctly reconstructed. The
cluster deconvolution is geometrically local, and thus trivially
parallel, and can be performed in parallel on the raw data.

\subsubsection{Hough Transform}

The Hough transform is a standard tool in image analysis that allows
recognition of global patterns in an image space by recognition of
local patterns (ideally a point) in a transformed parameter space. The
basic idea is to find curves that can be parametrized in a suitable
parameter space. In its original form one determines a curve in
parameter space for a signal corresponding to all possible tracks with
a given parametric form to which it could possibly belong~\cite{hough59}. 
All such curves belonging to the different signals are drawn in parameter
space. That space is then discretized and entries are stored in a
histogram. If the peaks in the histogram exceeds a given threshold,
the corresponding parameters are found.  

As mentioned above, in ALICE the local track model is a helix. In
order to simplify the transformation, the detector is divided into
subvolumes in pseudo-rapidity. If one restricts the analysis to tracks
originating from the vertex, the circular track in the $\eta$-volume
is characterized by two parameters: the emission angle with the beam
axis, $\psi$ and the curvature $\kappa$. The transformation is
performed from (R,$\phi$)-space to ($\psi$,$\kappa$)-space using the
following equations: 
%\begin{equation*}
%R = \sqrt{x^2+y^2}
%\end{equation*}
%\begin{equation*}
%\phi = \arctan(\frac{y}{x})
%\end{equation*}
%\begin{equation}
%\kappa = \frac{2}{R}\sin(\phi - \psi)
%\end{equation}
\begin{eqnarray}
R &=&\sqrt{x^2+y^2} \nonumber \\
\phi &=& \arctan(\frac{y}{x}) \nonumber \\
\kappa &=& \frac{2}{R}\sin(\phi - \psi) \nonumber \\
\end{eqnarray}

Each ADC value above a certain threshold transforms into a sinusoidal
line extending over the whole $\psi$-range of the parameter space. All
the corresponding bins in the histogram are incremented with the
corresponding ADC value. The superposition of these point
transformations produces a maximum at the circle parameters of the
track. The track recognition is now done by searching for local maxima
in the parameter space. 

\begin{figure}[hbt]
\centering
\includegraphics[height=8cm,width=4cm,angle=-90]{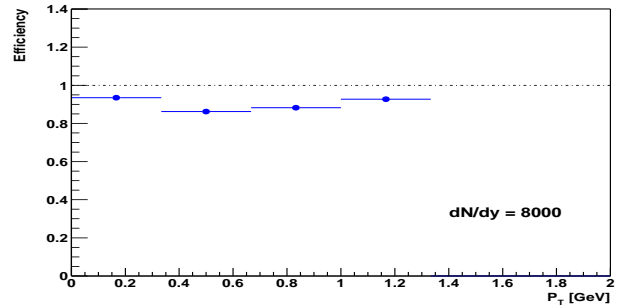}
\caption{Tracking efficiency for the Hough transform on a high occupancy event. 
The overall efficiency is above 90\%.}
\label{hougheff8000}
\end{figure}

Fig.\,\ref{hougheff8000} shows the tracking efficiency
for the Hough transform applied on a full multiplicity event and a
magnetic field of 0.2T. An overall efficiency above 90\% was
achieved. The tracking efficiency was taken as the
number of verified track candidates divided with the number of
generated tracks within the TPC acceptance. The list of verified track
candidates was obtained by taking the list of found local maxima and
laying a road in the raw data corresponding to the track parameters of
the peak. If enough clusters were found along the road, the track
candidate was considered a track, if not the track candidate was
disregarded. 

However, one of the problems encountered with the Hough transform
algorithm is the number of fake tracks coming from spurious peaks in the parameter
space. Before the tracks are verified by looking into the raw data, the
number of fake tracks is currently above 100\%. This problem has to be
solved in order for the tracks found by the Hough transform to be used
as an efficient input for the cluster fitting and deconvoluting
procedure\footnote{That is also reason, why the efficiency
does not drop for very low $p_T$ tracks like the offline
tracker in Fig.\,\ref{compresseff}. Strictly speaking, the efficiency
shown in Fig.\,\ref{hougheff8000} merely represents the quality
of the track candidates after the fakes have been removed.}.  
%Different solutions to the problem are currently being investigated. 

\subsubsection{Timing performance}

\begin{figure}[hbt]
\centering
\includegraphics[width=8cm]{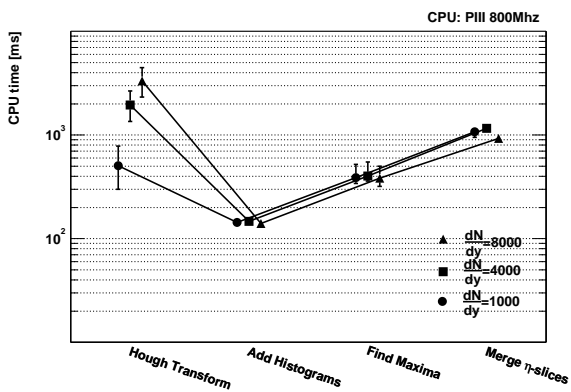}
\caption{Computation time measured on an 800\,MHz processor for different 
TPC occupancies and resolved with respect to the different processing 
steps for the Hough transform approach.}
\label{timinghough}
\end{figure}

Fig.\,\ref{timinghough} shows a timing measurement of the Hough based
algorithm for different particle multiplicities. The Hough transformation
is computed in parallel locally on each receiving node, whereas the other steps
(histogram adding, maxima finding and merging tracks across
$\eta$-slices) are done sequentially on the sector level. The
histograms from the different sub-sectors are added in order to
increase the signal-to-noise ratio of the peaks. For particle
multiplicities of dN/dy=8000, the four steps require about 700
seconds per event corresponding to 140,000 CPUs for a 200\,Hz event
processing rate. It should be noted  that the algorithm was already
optimized but some additional optimizations are still believed to be
possible. However, present studies indicate that one should not expect
to gain more than a factor of 2 without using hardware specifics of a
given processor architecture.  

The advantage of the Hough transform is that it has a very high degree
of locality and parallelism, allowing for the efficient use of FPGA
co-processors. Given the hierarchy of the TPC data analysis, it is
obvious that both the Hough transformation and the cluster
deconvolution can be performed in the receiver nodes. The Hough
transformation is particular I/O-bound as it create large histograms
that have to be searched for maxima, which scales poorly with modern
processor architectures and is ideally suited for FPGA
co-processors. Currently different ways of implementing the 
above outlined Hough transform in hardware are being investigated. 

%Although more research has to be undertaken in order to reduce the 
%fake track candidates and to implement the circuit in hardware.

% $Id: compression.tex,v 1.8 2003/10/06 17:16:25 loizides Exp $

\section{Data modeling and Data compression}

One of the mains goals of the HLT is to compress data efficiently with a
minimal loss of physics information.

In general two modes of data compression can be considered: 
\begin{itemize}
\item {\bf Binary lossless data compression}, allowing bit-by-bit
reconstruction of the original data set.
\item {\bf Binary lossy data compression}, not allowing bit-by-bit
reconstruction of the original data, while retaining however all
relevant physical information.
\end{itemize}

Methods such as Run-length encoding (RLE), Huffman and LZW are 
considered lossless compression, while thresholding and hit finding 
operations are considered lossy techniques that could lead to a loss of small
clusters or tail of clusters. It should be noted that data compression
techniques in this context should be considered lossless 
from a physics point of view. Many of the state of the art 
compression techniques were studied on
simulated TPC data and presented in detail in~\cite{berger02}. They
all result in compression factors of close to 2. However, the most
effective data compression can be done by cluster and track
modeling, as will be outlined in the following.

\subsection{Cluster and track modeling}
From a data compression point of the view, the aim of the track
finding is not to extract physics information, but to build a data
model, which will be used to collect clusters and to code cluster
information efficiently. Therefore, the pattern recognition algorithms
are optimized differently, or even different methods can be used
compared to the normal tracking.

The tracking analysis comprises of two main steps: Cluster
reconstruction and track finding. Depending on the occupancy, the
space points can be determined by a simple cluster finding or require
more complex cluster deconvolution functionality in areas of high
occupancy (see \ref{seqtracker} and \ref{ittracking}). 
In the latter case a minimum track model may be required in
order to properly decode the digitized charge clouds into their
correct space points. 

In any case the analysis process is two-fold: clustering and
tracking. Optionally the first step can be performed online while
leaving the tracking to offline, and thereby only recording the space
points. Given the high resolution of space points on one hand, and the
size of the chamber on the other, would result in rather large
encoding sizes for these clusters. However, taking a preliminary
zeroth order tracking into account, the space points can be encoded
with respect to their distance to such tracklets, leaving only small
numbers which can be encoded very efficiently. The quality of the
tracklet itself, with the helix parameters that would also be
recorded, is only secondary as the tracking is repeated offline with
the original cluster positions.  

\begin{table}
\begin{center}
\caption{Track parameters and their respective size}
\label{trackparams}
\begin{tabular}{|c|c|}
\hline
{\bf Track parameters}      & {\bf Size (Byte)} \\
\hline \hline
   Curvature               &       4 (float)\\
   X$_0$,Y$_0$,Z$_0$       &       4 (float)\\
   Dip angle,              &       4 (float)\\
   Azimuthal angle         &       4 (float)\\
   Track length            &       2 (integer)\\
\hline
\end {tabular}
\end{center}
\end{table}

\begin{table}
\begin{center}
\caption{Cluster parameters and their respective size}
\label{clusterparams}
\begin{tabular}{|c|c|}
\hline
{\bf Cluster parameters}    &        {\bf Size (Bit)} \\
\hline \hline
   Cluster present         &                1\\
   Pad residual            &                9\\
   Time residual           &                9\\
   Cluster charge          &                13\\
\hline
\end {tabular}
\end{center}
\end{table}

\subsection{Data compression scheme}

The input to the compression algorithm is a lists of tracks and their
corresponding clusters. For every assigned cluster, the
cluster centroid deviation from the track model is calculated in both
pad and time direction. Its size is quantized with respect to the
given detector resolution\footnote{The quantization steps have been
set to 0.5\,mm for the pad direction and 0.8\,mm for the time direction, 
which is compatible with the intrinsic detector resolution.}, and
represented by a fixed number of bits. In addition the 
total charge of the cluster is stored.
Since the cluster shape itself can be parametrized as a
function of track parameters and detector specific parameters, the
cluster widths in pad and time are not stored for every
cluster. During the decompression step, the cluster centroids are
restored, and the cluster shape is calculated based on the track
parameters. In tables~\ref{trackparams} and~\ref{clusterparams}, 
the track and cluster parameters are listed together with 
their respective size being used in the compression. Instead of
assigning only found clusters and their padrow numbers to a track,
we store for every padrow a cluster structure with a minimum size
of one bit, indicating whether the cluster is ``present'' or not.

\begin{figure}[thb]
\centering
\includegraphics[width=7cm]{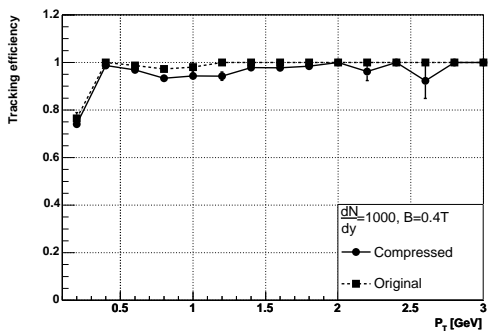}
\caption{Comparison of the tracking efficiency of the offline
reconstruction chain before and after data compression. A total loss
of efficiency of $\sim$1\% was observed.}
\label{compresseff}
\end{figure}

\begin{figure}[thb]
\centering
\includegraphics[width=7cm]{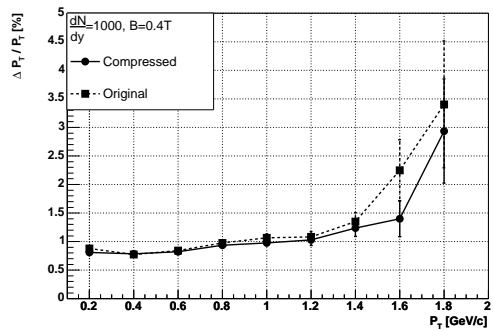}
\caption{Comparison of the $p_T$ resolution of the offline reconstruction
chain before and after data compression.}
\label{compressres}
\end{figure}

The compression scheme has been applied to a simulated PbPb 
event with a multiplicity of $dN/dy=1000$. The input tracks used for the
compression are tracks reconstructed with the sequential 
tracking approach. The {\it remaining clusters}, or the clusters
which were not assigned to any tracks during the track finding step,
were disregarded and not stored for further analysis\footnote{The
remaining clusters mainly originate from very low $p_T$ tracks such
as $\delta$-electrons, which could not be reconstructed by the track
finder. Their uncompressed raw data amounts to a relative size of about 20\%.}. 
A relative size of 11\% for the compressed data with respect 
to the original set is obtained. 

In order to evaluate the impact on the physics observables, 
the compressed data is decompressed and the restored cluster 
are processed by the offline reconstruction chain. 
In Fig.~\ref{compresseff} the offline tracking efficiency before 
and after applying the compression is compared as a function 
of $p_T$. A total loss of about 2\% in efficiency is observed.
Fig.\,\ref{compressres} shows for the same events the $p_T$ resolution 
as a function of $p_T$ before and after the compression is applied.
The observed improvement of the $p_T$ resolution is connected
to way the errors of the cluster are calculated. For the case of
the standard offline reconstruction chain the errors are calculated
using the cluster information itself, whereas for the compression scheme
they are calculated using the track parameters. 

Keeping the potential gain of statistics by the increased 
event rate written to tape in mind, one has to weight 
the tradeoff between the impact on the physics observables 
and the cost for the data storage. For occupancy events of
more than 20\% (corresponding to $dN/dy>2000$), 
clusters start to overlap and has to be properly deconvoluted 
in order to effectively compress the data. 

In this scenario, the Hough transform or another effective iterative
tracking procedure would serve as an input for the cluster
fitting/deconvolution algorithm. With a high online tracking
performance, track and cluster modeling, together with noise removal,
can reduce the data size by a factor of 10.

%\begin{figure}
%\centering
%\includegraphics[width=2.5in]{myfigure}
% where an .eps filename suffix will be assumed under latex, 
% and a .pdf suffix will be assumed for pdflatex
%\caption{Simulation Results}
%\label{fig_sim}
%\end{figure}

%\begin{figure*}
%\centerline{\subfigure[Case I]{\includegraphics[width=2.5in]{subfigcase1}
% where an .eps filename suffix will be assumed under latex, 
% and a .pdf suffix will be assumed for pdflatex
%\hfil
%\subfigure[Case II]{\includegraphics[width=2.5in]{subfigcase2}
% where an .eps filename suffix will be assumed under latex, 
% and a .pdf suffix will be assumed for pdflatex
%\label{fig_second_case}}}
%\caption{Simulation results}
%\label{fig_sim}
%\end{figure*}

%\begin{table}
%% increase table row spacing, adjust to taste
%\renewcommand{\arraystretch}{1.3}
%\caption{An Example of a Table}
%\label{table_example}
%\centering
%% Some packages, such as MDW tools, offer better commands for making tables
%% than the plain LaTeX2e tabular which is used here.
%\begin{tabular}{|c||c|}
%\hline
%One & Two\\
%\hline
%Three & Four\\
%\hline
%\end{tabular}
%\end{table}

% $Id: conclusions.tex,v 1.6 2003/10/06 17:16:25 loizides Exp $

\section{Conclusion}

Focusing on the TPC, the sequential approach, which consists of cluster
finding followed by track finding, is applicable for pp and low
multiplicity PbPb data up to dN/dy of 2000 to 3000 with more than
90\% efficiency. The timing results indicate that the desired
frequency of 1KHz for pp and 200 Hz for PbPb can be achieved. For
higher multiplicities of dN/dy $\ge$ 4000 the iterative approach using
the Circle Hough transform for primary track candidate finding shows
promising efficiencies of around 90\% but with high computational
costs. 

By compressing the data using data modeling techniques, the results for low
multiplicity events show that one can compress data of up to 10\% 
relative to the original data sizes with a small loss of the tracking 
efficiency of about 2\%, but slightly improved $p_T$ resolution. 
%However, keeping the potential gain of statistics by
%the increased event rate written to tape in mind, one has to weigh the tradeoff
%between the impacts on the physical observables and the costs for the
%data storage. 

%\section*{Acknowledgment}
% optional entry into table of contents (if used)
%\addcontentsline{toc}{section}{Acknowledgment}

%\IEEEtriggeratref{10}

\end{document}